\documentclass[journal]{IEEEtran}
\IEEEoverridecommandlockouts
\usepackage{ifpdf}
\usepackage{cite}
\usepackage[pdftex]{graphicx}
\usepackage{amsmath,amssymb,amsfonts}
\usepackage{algorithmic}
\usepackage{array}
\usepackage[caption=false,font=footnotesize]{subfig}
\usepackage{fixltx2e}
\usepackage{stfloats}
\usepackage{textcomp}
\usepackage{breqn}
\usepackage{mwe}
\usepackage{stackrel}
\usepackage{enumitem}
\usepackage{gensymb}
\usepackage{color}
\usepackage[dvipsnames]{xcolor}
\usepackage{balance}
\usepackage{multirow}
\usepackage{soul}
\usepackage{tabto}
\usepackage{txfonts}
\usepackage{hyperref}
\usepackage{paralist}
\usepackage{url}
\ifCLASSOPTIONcompsoc
    \usepackage[caption=false, font=normalsize, labelfont=sf, textfont=sf]{subfig}
\else
\usepackage[caption=false, font=footnotesize]{subfig}
\fi
\def\BibTeX{{\rm B\kern-.05em{\sc i\kern-.025em b}\kern-.08em
    T\kern-.1667em\lower.7ex\hbox{E}\kern-.125emX}}
\begin{document}

\title{High-Fidelity Model of Stand-Alone Diesel Electric Generator with Hybrid Turbine-Governor Configuration for Microgrid Studies} 

\author{Chinmay Shah,~\IEEEmembership{Student Member,~IEEE,}
        Mariko Shirazi,~\IEEEmembership{Member,~IEEE,}
        Richard Wies,~\IEEEmembership{Senior Member,~IEEE,}
        Phylicia Cicilio,~\IEEEmembership{Member,~IEEE,}
        Timothy Hansen,~\IEEEmembership{Senior Member,~IEEE,},
        and Reinaldo Tonkoski,~\IEEEmembership{Senior Member,~IEEE}
        
\thanks{This work is supported by the U.S. Department of Energy Office of Science, Office of Basic Energy Sciences, EPSCoR Program; Office of Electricity, Microgrid R\&D Program; and Office of Energy Efficiency and Renewable Energy, Solar Energy Technology Office under the EPSCoR grant number DE-SC0020281.}
\thanks{Chinmay Shah is with the Department
of Electrical and Computer Engineering, University of Alaska Fairbanks, Fairbanks,
AK, 99709 USA e-mail: cshah@alaska.edu.}
\thanks{Phylicia Cicilio is with Alaska Center for Energy and Power, University of Alaska Fairbanks, Fairbanks,
AK, 99709 USA.}
\thanks{Mariko Shirazi is with Alaska Center for Energy and Power, University of Alaska Fairbanks, Fairbanks,
AK, 99709 USA.}
\thanks{Richard Wies is with the Department
of Electrical and Computer Engineering, University of Alaska Fairbanks, Fairbanks,
AK, 99709 USA.}
\thanks{Timothy Hansen is with the Department
of Electrical Engineering and Computer Science, South Dakota State University, Brookings,
SD, 57007 USA.}
\thanks{Reinaldo Tonkoski is with the Department
of Electrical Engineering and Computer Science, South Dakota State University, Brookings,
SD, 57007 USA.}
}

\maketitle

\begin{abstract}
Diesel electric generators are an inherent part of remote hybrid microgrids found in remote regions of the world that provide primary frequency response (PFR) to restore system frequency during load or generation changes. However, with inverter-based resources (IBR) integration into microgrids, the IBR control provides a fast frequency response (FFR) to restore the system frequency. Hence, supplementing PFR with FFR requires a sophisticated control system and a high fidelity diesel electric generator model to design these control systems. In this work, a high-fidelity model of a diesel electric generator is developed. Its parameters are tuned using a surrogate optimization algorithm by emulating its response during a load change to a 400 kVA Caterpillar C-15 diesel generator, similar to those found in remote microgrids. The diesel electric generator model consists of a synchronous machine, DC4B excitation with V/Hz limiter, and a proposed modified IEEE GGOV1 engine-governor model (GGOV1D). The performance of the GGOV1D is compared with simple, Woodward DEGOV, and a standard IEEE GGOV1 engine-governor model. Results show that error in the diesel electric generator's response to load changes using the GGOV1D model is lower with an improved frequency response during the arresting and rebound period than the other engine-governor models.
\end{abstract}


\section{Introduction}
\IEEEPARstart{R}{emote} islanded microgrids like those found in the Arctic and island nations are primarily powered by diesel electric generators, with increasing contributions from inverter-based renewable resources like wind and solar photovoltaics (PV). There is significant interest in adopting more renewable energy resources with the primary motivation of displacing imported fossil fuels to reduce the cost of electricity and heat \cite{Holdmann}. The integration of inverter-based generators (IBR) reduces the system's inertia. IBRs can cause numerous stability issues such as a high rate of change of frequency (ROCOF) due to low inertia \cite{Markovic, Tamrakar_VI_MDPI, Holttinen} and low-frequency oscillations introduced by the phase-locked loop. Detailed descriptions about these stability issues are presented in \cite{IEEEStabilityDef_2020,Farrokhabadi,Holttinen, san2020}. Also, accurate models of such hybrid systems are necessary to analyze the system's stability. While there has been much focus on high-fidelity IBR models, it is just as important to have a high-fidelity diesel electric generator model. The existing diesel electric generator models available in the literature are for small capacity diesel electric generators and not for hundreds of kilowatts hybrid microgrid scale diesel electric generators. Also, it was found that the existing diesel engine-governor models typically employed in the literature produce poor frequency response compared to our experimental data. Hence, the main focus of this work is to develop a high-fidelity model of the diesel electric generator adapting the engine-governor model with improved frequency response during the arresting and rebound period that can be integrated with IBR models to study the inherent stability issues. 
\begin{figure}[!htp]
    \centering
    \includegraphics[width=0.65\columnwidth]{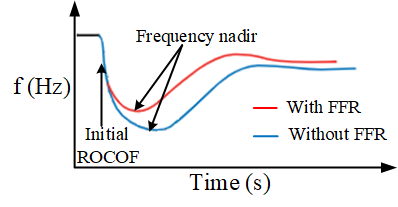}
    \caption{Comparison of grid frequency during load change with and without fast frequency response (FFR) \cite{SU2020}.}
    \label{fig:freq_response_comp}
\end{figure}

Historically, diesel electric generators have provided primary frequency response (PFR) during load changes in diesel-only microgrids. However, microgrids with diesel electric generators and IBRs have high ROCOF and can be improved by fast frequency response (FFR) provided by the frequency response control system of an IBR, as shown in Fig.~\ref{fig:freq_response_comp}. The FFR is defined as the contribution of electric power from the IBRs or energy storage that rapidly responds to the frequency changes to minimize the torque imbalance of the synchronous generators \cite{SU2020, matevosyan2014}. Hence FFR can supplement PFR to prevent an initial rapid decline in microgrid frequency during the arresting period and withdraw after the frequency nadir when PFR is sufficient for restoring the frequency. However, this requires a coordinated control between the diesel governor action and the IBR controller. A high-fidelity diesel electric generator model with accurate governor action is needed to design coordinated control systems as remote islanded microgrids continue to integrate more IBRs.

A high-fidelity diesel electric generator model is developed in this work with complete transient models of the electric machine, exciter, engine-governor, and other limiting components. The electric machine of a diesel electric generator is commonly modeled using standard synchronous machine models \cite{Kundur}, often simulated in MATLAB\textsuperscript{\textregistered} Simulink\textsuperscript{\textregistered} with a sixth-order state-space model, accounting for the stator, field, and damper winding dynamics \cite{MatlabSyn}. Numerous studies have used this model for transient simulations and have reported their model parameter values~\cite{Nikkhajoei, Yeager, AbuHussein, Sakamoto, Gish, Mirosevic, Eberlein, Leza}. Now, numerous dynamic simulations of diesel electric generators have implemented simplified diesel genset models, often containing only one or two first-order transfer functions to simulate the complete diesel electric generator system \cite{Papathanassiou, Datta, Long}. These simplified models neglect significant dynamics, such as stator, field, damper winding, saturation, and diesel engine and fuel flow dynamics. Also, the diesel electric generator considered in \cite{Long} had a capacity of 16 kVA. 
Diesel electric generators operating in remote islanded microgrids in Alaska or other Arctic regions and island nations are on the scale of hundreds of kVA. Their behavior is much different from the 16 kVA diesel electric generator, so the models do not translate or scale well for higher power capacity generators.

The diesel electric generator model developed in this work uses a 320 kW/400 kVA synchronous machine, a DC4B excitation system with a V/Hz limiter as presented in previous work \cite{shahdiesel2021}, and a proposed engine-governor model that emulates a Caterpillar C-15 (CAT C-15) diesel engine, and an electronic governor model with unknown structure. The frequency response of the diesel electric generator using the standard simple engine-governor model \cite{stavrakakis1995}, Woodward DEGOV engine-governor model \cite{DEGOV}, and  IEEE GGOV1 engine-governor model\cite{neplangov} did not accurately emulate the actual diesel electric generator frequency response. Hence, this work proposed a modified IEEE GGOV1 engine-governor model (GGOV1D), and its performance was compared with the standard engine-governor models. The IEEE GGOV1 governor model in the literature is developed to model gas or steam turbines, but it can be used for a diesel electric generator with an electronic governor \cite{neplangov}. The IEEE GGOV1 model engine gain value and no-load fuel flow were calculated using the experimental data to make it compatible with the diesel electric generator model and better emulate the frequency response. 

Several optimization algorithms are proposed in the literature to tune the parameters of the diesel electric generator components. The whale optimization heuristic algorithm was used in \cite{huang2020}; however, the authors of this work have only tuned the parameters based on active and reactive power response and neglected the voltage and frequency response. The Box-Constrained Levenberg-Marquardt algorithm was used by the authors of \cite{Long} that fits the frequency and voltage responses, but neglects the fitting of active and reactive power response of the diesel electric generator. A limitation of the Levenberg-Marquardt algorithm is that it requires a reasonable initial estimate of the parameters, and failure to do so affects the algorithm's convergence to optimal values. The authors of \cite{Long} used a genetic algorithm to generate reasonable initial estimates for the Levenberg-Marquardt optimization algorithm. Using two stage optimization algorithms makes the parameter identification process computationally intensive. A radial basis function surrogate model optimization algorithm was used to tune the parameters of the diesel electric generator transient models \cite{Khazeiynasab2021, matlabsopt2021}. The surrogate optimization searches for the best value by evaluating its surrogate on thousands of points and inputs the best approximation to the objective function to minimize the error. Since the surrogate optimization evaluates thousands of points at one time, it takes less time to evaluate and obtain the global minimum \cite{matlabsurrogate}. The surrogate optimization algorithm was used to tune the parameters of the diesel electric generator in this work.

The main contribution of this work is the proposed experimentally validated modified IEEE GGOV1 engine-governor model for diesel electric generator applications (GGOV1D). The GGOV1D model's performance is compared against the other engine-governor models. The surrogate optimization algorithm is used to identify optimal parameters of the engine-governor models based on the active power, reactive power, voltage, and frequency experimental responses of the 320 kW/400 kVA CAT C15 diesel electric generator.

The rest of the article is organized as follows: Section~\ref{section2} presents the modeling of system components of the diesel electric generator. Section~\ref{section3} provides a brief description of the surrogate optimization algorithm. This is followed by an overview of the CAT C15 diesel electric generator, experimental setup, and data acquisition in Section~\ref{section4}. The simulation results are presented in Section~\ref{section5}, followed by the conclusion in Section~\ref{section6}.

\begin{figure}[!htp]
    \centering
    \includegraphics[width=0.75\columnwidth]{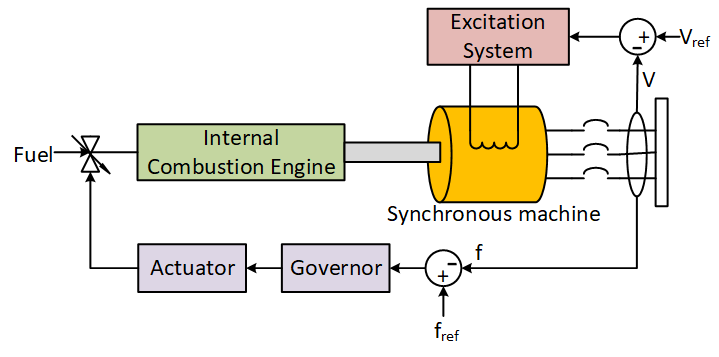}
    \caption{Diesel electric generator block diagram.}
    \label{fig:diesel_gen}
\end{figure}
\section{Diesel electric generator Dynamic Model}\label{section2}
The block diagram of the diesel electric generator is comprised of a synchronous machine, excitation system, speed governor, actuator, and internal combustion diesel engine as shown in Fig.~\ref{fig:diesel_gen}.
The excitation system provides direct current (DC) to the rotor winding in the synchronous generator. The diesel engine spins the rotor, creating a rotating magnetic field that induces emf in the stator winding. The speed governor controls the fuel flow to the diesel engine by actuating the valve, which controls the rotor's speed.

\subsection{Synchronous Machine Model}
\begin{figure}[!htp]
    \centering
  \subfloat[\label{fig:syncmach_daxis}]{%
       \includegraphics[width=0.75\columnwidth]{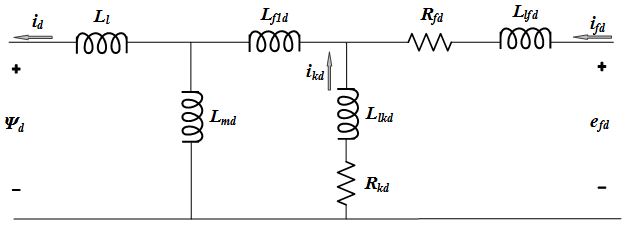}}
    \vfill
  \subfloat[\label{fid:syncmach_qaxis}]{%
        \includegraphics[width=0.5\columnwidth]{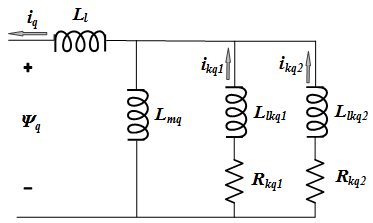}}
  \caption{(a) Per-unit d-axis equivalent circuit of synchronous machine, (b) Per-unit q-axis equivalent circuit of synchronous machine \cite{ieee1110-2019}.}
  \label{fig:syncmach} 
\end{figure}

The synchronous machine model used in this work is adopted from the IEEE 1110-2019 standard \cite{ieee1110-2019} and the direct and quadrature axis equivalent circuit diagram is shown in Fig.~\ref{fig:syncmach}.
The d-axis and q-axis armature and field voltage and the flux linkage from the equivalent diagram in Fig.~\ref{fig:syncmach} is given by:
\begin{align}
    \begin{split}\small
        \begin{bmatrix} \Psi_{d} \\ \Psi_{kd}\\ \Psi_{fd} \end{bmatrix} = \begin{bmatrix}
            L_{md}+L_{l} & L_{md} & L_{md}\\ L_{md} & L_{lkd}+L_{f1d}+L_{md} & L_{f1d}+L_{md} \\
            L_{md} & L_{f1d}+L_{md} & L_{lfd}+L_{f1d}+L_{md} \end{bmatrix} \begin{bmatrix}
                -i_{d} \\ i_{kd}\\ i_{fd} \end{bmatrix}
        \label{eq:1}
    \end{split} \\
    \begin{split}\small
        \begin{bmatrix} \Psi_{q} \\ \Psi_{kq1}\\ \Psi_{kq2} \end{bmatrix} = \begin{bmatrix}
            L_{mq}+L_{l} & L_{mq} & L_{mq}\\ L_{mq} & L_{mq}+L_{kq1} & L_{mq} \\
            L_{md} & L_{f1d}+L_{md} & L_{lfd}+L_{f1d}+L_{md} \end{bmatrix} \begin{bmatrix}
                -i_{q} \\ i_{kq1}\\ i_{kq2} \end{bmatrix}
        \label{eq:2}
    \end{split}
\end{align}
\begin{equation}
    V_d = -i_{d}R_{s} - \omega\Psi_{q} + \frac{d\Psi_{d}}{dt}
    \label{eq:3a}
\end{equation}
\begin{equation}
    V_q = -i_{q}R_{s} - \omega\Psi_{d} + \frac{d\Psi_{q}}{dt}
    \label{eq:3b}
\end{equation}
\begin{equation}
    V_0 = -i_{0}R_{0} + \frac{d\Psi_{0}}{dt}
    \label{eq:4}
\end{equation}
\begin{equation}
    V_{fd} = \frac{d\Psi_{fd}}{dt} + R_{fd}i_{fd}
    \label{eq:5}
\end{equation}
where $s$ is the stator quantity, $l$ and $m$ are the leakage and magnetizing inductance, respectively, and $f$ and $k$ are the field and damper winding quantities, respectively. The operational transient and sub-transient reactance and time constants can be obtained from the equivalent circuit presented above using the data translation approach described in the IEEE 1110-2019 standard. However, it is recommended to use the measured quantities instead. Hence in this work the transient and sub-transient reactance and open circuit time constant values provided in the diesel electric generator technical specification sheet are used.

\subsection{Excitation System Model}
\begin{figure}[h]
    \centering
	\includegraphics[width=0.99\columnwidth]{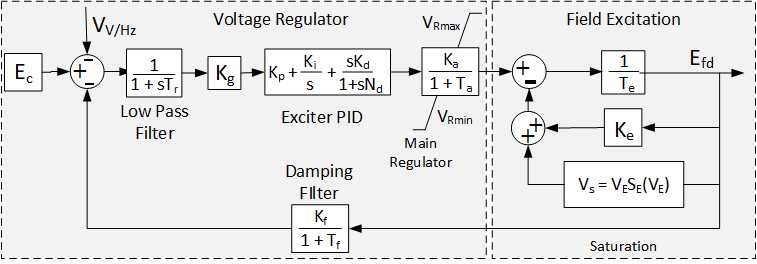}
    \caption{DC4B exciter block diagram as implemented in \cite{DC4B}}
    \label{fig:DC4B}
\end{figure}

The excitation system for the diesel electric generator is the system that provides the field current to the rotor winding. The DC4B exciter model \cite{DC4B}, as illustrated in Fig.~\ref{fig:DC4B}, emulates the excitation system presented in the datasheet for the CAT C15 diesel electric generator used in this work. The DC4B excitation system utilizes a field-controlled DC commutator exciter with a continuously acting voltage regulator that obtains input from the generator terminals. The DC4B excitation system also has the following parameter restrictions \cite{SMexcitation}:
\begin{inparaenum}
    \item The stabilization feedback time constant $T_f$ can only be set to zero if the feedback gain $K_f$ equals zero.
    \item The stabilization feedback loop is included in the model only if the PID controller's derivative action is not used.
    \item Efd$_1$ $>$ Efd$_2$ and SeEfd$_1$ $>$ SeEfd$_2$
\end{inparaenum}
where Efd$_1$ and Efd$_2$ are the exciter voltages at which the exciter saturation is defined, and SeEfd$_1$ and SeEfd$_2$ are the exciter saturation function values.
The variations in the input voltage are compensated by the system loop gain $k_g$, along with the PID controller. With fixed PID gain values, the loop gain $K_g$ determines the position of the poles of the generator-exciter-controller closed loop system as explained in \cite{kim2005}. The location of the system poles and zeros determine the performance of the excitation control system. Based on the responses presented in \cite{kim2005} for different values of $K_g$, it is an extremely sensitive parameter that impacts the overall response of the diesel electric generator. For this work a range for $K_g$ between 0 and 1 is chosen for tuning and optimization. 

\begin{figure}[!htp]
    \centering
	\includegraphics[width=0.75\columnwidth]{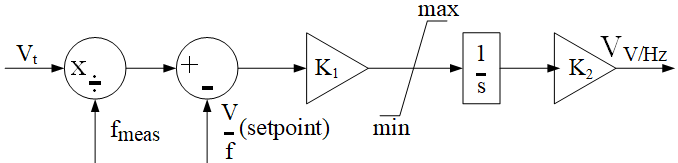}
    \caption{V/Hz limiter.}
    \label{fig:voltshertz}
\end{figure}

In this work, a V/Hz limiter is added to the standard DC4B exciter model to prevent over-fluxing due to over-voltage or under-frequency conditions that could damage the machine. The implementation of the V/Hz limiter is shown in Fig.~\ref{fig:voltshertz} and is based on a hydroelectric generator V/Hz limiter implementation \cite{Maina}. The V/Hz limiter takes the terminal voltage and the measured frequency ratio and compares it to a setpoint ratio. If the error between the voltage and frequency ratio and the setpoint is positive, then the integrator acts to produce a negative signal that reduces the voltage setpoint sent to the exciter until the V/Hz ratio is below the setpoint. The integrator is reset as soon as the error becomes negative, and the limiting signal sent to the exciter is zero.

\subsection{Engine-Governor Models}
The purpose of the governor in a diesel electric generator is to control its speed by controlling the flow of fuel to the engine cylinders. Governors can be categorized into mechanical governors, electronic governors, and electronic injection governors. This work considers an electronic governor consisting of three parts: the sensing element, the controller, and the actuator. 

However, the parameters and the model structure of the governor are unknown. Hence, four engine-governor dynamic models are used in this work to evaluate which model provides the most accurate response. Those engine-governor models are as follows: 1) Simple engine-governor model, 2) Woodward DEGOV engine-governor model, 3) IEEE GGOV1 turbine/engine-governor model, and 4) GGOV1D engine-governor model. The DEGOV engine-governor model is the traditional diesel governor model commonly used in transient modeling of diesel electric generators \cite{DEGOV}.

\subsubsection{Simple Engine-Governor Model}
\begin{figure}[!htp]
    \centering
    \includegraphics[width=\columnwidth]{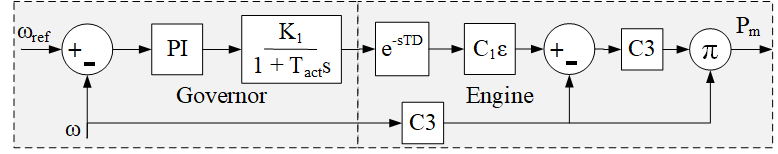}
    \caption{Simple engine-governor block diagram for diesel electric generator operation in isochronous mode.}
    \label{fig:simplegov}
\end{figure}

The simple engine-governor model introduced in this paper is a coalescence of a diesel engine model, speed regulator, and an actuator and is adopted from \cite{stavrakakis1995}. The mechanical power output $P_m$ of the diesel engine is given by
\begin{equation}
    P_m = C_2\omega(p_i-p_f)
    \label{eq:7}
\end{equation}
\begin{equation}
    p_i = C_1m_b^{'}\varepsilon
    \label{eq:8}
\end{equation}
\begin{equation}
    p_f = C_3\omega
    \label{eq:9}
\end{equation}
where $C_1$, $C_2$, and $C_3$ are the proportionality constants, $p_i$ is the effective pressure of the diesel engine, $p_f$ is the mean pressure of the mechanical losses, $m_b^{'}$ is the fuel consumption rate (lit/hr), $\omega$ is the speed of the rotor (rad/s), and $\varepsilon$ is the combustion efficiency of the diesel engine. The speed of the rotor is controlled by the diesel engine and is dependent on the fuel intake. The differential equation defining the rate of change of fuel consumption and the speed regulation in isochronous mode of the diesel electric generator is given by
\begin{equation}
    \frac{dm_{b}}{dt} = \frac{1}{t_{act}}\Bigg(-K_{1}\int(\frac{K_{I}}{\omega_{ref}}\Delta\omega) - \frac{K_1K_P}{\omega_{ref}}\Delta\omega - m_b \Bigg)
    \label{eq:10}
\end{equation}
\begin{equation}
    m_b^{'}(t) = m_b(t-\tau_d)
    \label{eq:11}
\end{equation}
where $K_1$ is the actuator gain, $K_P$ and $K_I$ are the proportional and integral controller gains, $\Delta\omega$ is the speed error, and $\tau_d$ is the engine delay. The block diagram of the implemented simple engine-governor model based on eq.~\eqref{eq:7}, \eqref{eq:10}, and \eqref{eq:11} is shown in Fig.~\ref{fig:simplegov}.

\subsubsection{DEGOV Engine-Governor Model}
\begin{figure}[!htp]
    \centering
	\includegraphics[width=\columnwidth]{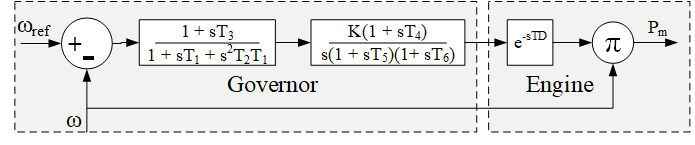}
    \caption{Woodward diesel engine-governor DEGOV block diagram.}
    \label{fig:DEGOV}
\end{figure}

The Woodward diesel engine-governor DEGOV model \cite{DEGOV} is the same as presented in \cite{pworld, neplangov} and illustrated in Fig.~\ref{fig:DEGOV}. The variables $T_1, T_2, T_3, T_4, T_5,$ and $T_6$ are the actuator time constants. $T_D$ is the engine delay and $\omega$ and $\omega_{ref}$ are the rotor and reference speed.

\subsubsection{IEEE GGOV1 Model}
The IEEE GGOV1 is commonly used to model the turbine-governor dynamics of gas turbines in large electric grids \cite{pestr12013}. However, the model is also suitable for diesel electric generators with modern electronic or digital governors \cite{neplangov}. The GGOV1 model has the following components:
\begin{inparaenum}
    \item Turbine dynamics that represent the turbine/engine model, fuel flow, and valve position limits, 
    \item Speed control,
    \item Temperature control,
    \item Acceleration control, and
    \item Load control.
\end{inparaenum}
All of these control modes compete for the overall control of the system by regulating the main fuel valve via a ``low value select block.'' The model also can operate in isochronous or droop control mode via a ``Rselect'' block. Along with different control modes, the IEEE GGOV1 model has the following advantages as compared to the other engine-governor models: valve position limits and calculation and fuel flow dynamics. 

\begin{figure}[!htp]
    \centering
    \includegraphics[width=\columnwidth]{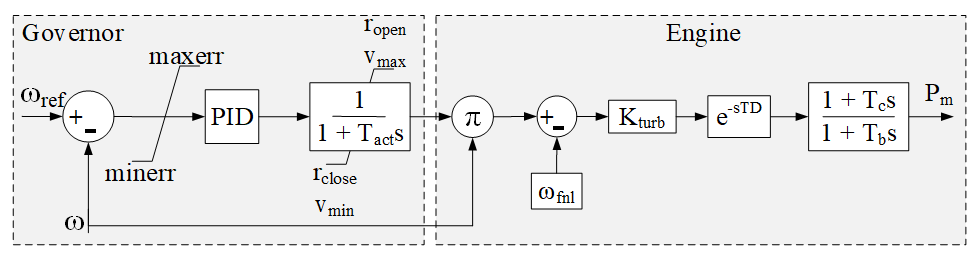}
    \caption{Modified IEEE GGOV1 block diagram.}
    \label{fig:ggov1}
\end{figure}

The CAT C-15 diesel electric generator's electronic control module (ECM) inputs the coolant temperature and intake manifold temperature to adjust the fuel rate for combustion and reduce the exhaust smoke. However, it does not impact the output of the generator. Hence, the temperature control loop is not considered in this work. The acceleration control mode is mainly used during the startup, which rarely functions during the regular grid operation and is also eliminated. The diesel electric generator tested in this work is stand-alone, operates in an isochronous mode, and therefore does not have a load control mode. The IEEE GGOV1 model with a speed control loop that is compatible with diesel electric generators is shown in Fig.~\ref{fig:ggov1} where, $\omega_{fnl}$ is no load fuel flow, and $T_b$ and $T_c$ are turbine/engine lag and lead time, respectively. The actuator transfer function in the IEEE GGOV1 model, as shown in Fig.~\ref{fig:ggov1} is a first-order model with no zeros and only one pole. The actuator model in the DEGOV engine-governor model is third-order with one pole at the origin, two dominant poles, and one zero. In this work, the IEEE GGOV1 model is further modified to have a third-order actuator transfer function.

\subsubsection{GGOV1D Model}
\begin{figure}[!htp]
    \centering
    \includegraphics[width=\columnwidth]{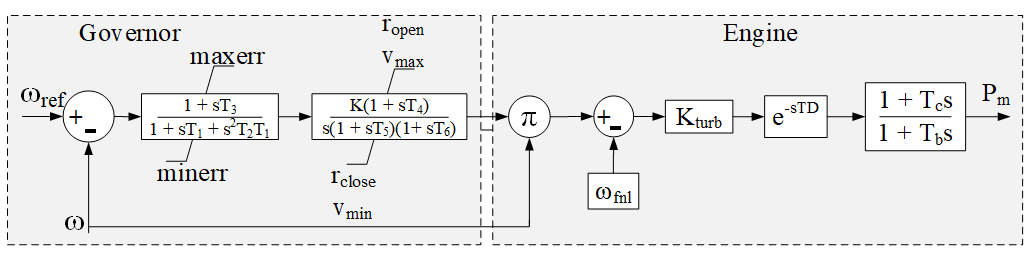}
    \caption{GGOV1D model with $3^{rd}$ order actuator.}
    \label{fig:ggov1_modified}
\end{figure}

The block diagram of the proposed GGOV1D model that has a third-order actuator transfer function is shown in Fig.~\ref{fig:ggov1_modified}. The two additional dominant poles in the model would increase the system's rise time, making the response sluggish. At the same time, the addition of the zero increases the overshoot and reduces the rise time and response time. Thus the additional zero and poles give us an extra degree-of-freedom to tune the governor model to obtain the desired response. Also, the response will be more stable if the zero is tuned to be closer to the dominant poles. The response of the proposed GGOV1D model and the standard IEEE GGOV1 model for the diesel electric generator is presented in the next section.

\begin{figure}[!htp]
    \centering
    \includegraphics[width=0.85\columnwidth]{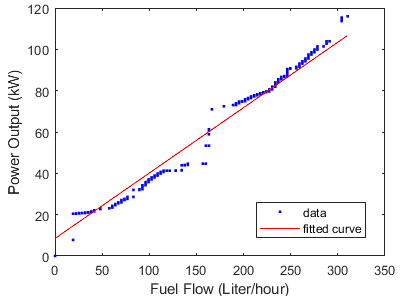}
    \caption{Fuel flow versus power output of CAT C15 diesel electric generator.}
    \label{fig:fuelflow_power}
\end{figure}

The parameter $K_{turb}$ is the turbine/engine gain which represents the mechanical model of the diesel engine and can be estimated from the plot of fuel flow versus power output. In this work, $K_{turb}$ is tabulated from the diesel fuel flow (liter/hour) versus diesel electric generator power output (kW) shown in Fig.~\ref{fig:fuelflow_power}. Considering the fitted curve in Fig.~\ref{fig:fuelflow_power} as a linear approximation of the relationship between the fuel flow and power output of the diesel electric generator, the tabulated value of $K_{turb}$ is $0.362$ and the no-load fuel flow $\omega_{fnl}$ is 0.12 pu. The base value to calculate no-load fuel flow in pu is the maximum fuel flow rate of the diesel electric generator.

\section{Surrogate Optimization Algorithm}\label{section3}
The general form of the optimization algorithm to estimate the parameters of a diesel electric generator is represented as follows:
\begin{align}
    \begin{split}
        \min\mathcal{G}(\Phi_{p})~~~~ \\
        s.t. ~\Phi_{p} \in [\Phi_{p}^{L}, \Phi_{p}^{U}]
    \end{split}
    \label{eq:14}
\end{align}
where $\Phi_{p}$ are the set of parameters to be estimated, and $\Phi_{p}^{L}$ and $\Phi_{p}^{U}$ are the lower and upper bounds of the diesel electric generator parameters $\Phi_{p}$, respectively. The parameters $\Phi_{p}$ can be estimated by minimizing the error between the measured and the simulated responses of $P$, $Q$, $V$, and $f$ tabulated in the previous section. The objective function that minimizes the error to estimate the parameters $\Phi_{p}$ when the analytical expression is unknown is represented as follows:
\begin{multline}
    \mathcal{G}(\mathcal{X},x) = w_p \times P_{nRMSE} + w_q \times Q_{nRMSE} + w_v \times V_{nRMSE}\\{} + w_f \times f_{nRMSE}
    \label{eq:15}
\end{multline}
where $w_p$, $w_q$, $w_v$, and $w_f$ are the weights, and $\mathcal{X} = [P_{meas}$, $Q_{meas}$, $V_{meas}$, $f_{meas}]$ are the active power, reactive power, voltage, and frequency, respectively, measured from the CAT C-15 diesel electric generator during a load change. $x = [P_{sim}$, $Q_{sim}$, $V_{sim}$, $f_{sim}]$ is the diesel electric generator simulation response during a load change. $P_{nRMSE}$, $Q_{nRMSE}$, $V_{nRMSE}$, and $f_{nRMSE}$ are the normalized root mean square errors tabulated as follows:
\begin{equation}
    P_{nRMSE} = \frac{1}{\Bar{p}}\sqrt{\frac{1}{N}\sum_{i=1}^N(P_{i,meas}-P_{i,sim})^2}
    \label{eq:16}
\end{equation}
where $\Bar{p}$ is the normalization factor. Similarly, nRMSE is tabulated for $Q$, $V$, and $f$.

Now, $n$ distinct random points $\Phi_{p}^{1}, \Phi_{p}^{2}, \hdots, \Phi_{p}^{n}$ are created within the bounds with the function values $\mathcal{G}^{1}, \mathcal{G}^{2}, \hdots, \mathcal{G}^{n}$ obtained to solve the problem defined in \eqref{eq:17} using surrogate optimization.  A surrogate of the objective function is then created by interpolating a radial basis function (RBF) as:
\begin{equation}
    s(\Phi_{p}) = \sum_{i=1}^{n}\lambda_i*\mathcal{B}(||\Phi_{p} - \Phi_{p}^{i}||) + \alpha(x)
    \label{eq:17}
\end{equation}
through the points $(\Phi_{p}^{1}, \mathcal{G}^{1})$, $(\Phi_{p}^{2}, \mathcal{G}^{2})$, $\hdots$, $(\Phi_{p}^{n}, \mathcal{G}^{n})$. The variable $\lambda_{i}$ is the weight and the norm $||.||$ is the Euclidean norm in $\mathbb{R}$. $\alpha$ is from $\Pi_{m}$, the space of polynomials of degree less than or equal to $m$ \cite{gutmann2001}. The $\mathcal{B}$ used in this work is cubic, i.e.,
\begin{equation}
    \mathcal{B}(r) = r^3
    \label{eq:18}
\end{equation}
where $r$ can be any variable. Once the surrogate value is obtained at $n$ random points, the minimum of the objective function is found by evaluating the following equation at point $i$:
\begin{equation}
    \mathcal{F}(\Phi_{p}^{i}) = w*\mathcal{S}(\Phi_{p}^{i}) + (1-w)*\mathcal{D}(\Phi_{p}^{i})
    \label{eq:19}
\end{equation}
where $w \in (0, 1)$ is the weight, $\mathcal{S}(\Phi_{p}^{i})$ is the scaled surrogate, and $\mathcal{D}(\Phi_{p}^{i})$ is the scaled distance. The scaled surrogate is defined by
\begin{equation}
    \mathcal{S}(\Phi_{p}^{i}) = \frac{s(\Phi_{p}^{i}) - s_{min}}{s_{max} - s_{min}}
    \label{eq:20}
\end{equation}
where $s_{min}$ is the minimum surrogate value among the $n$ random points, and $s_{max}$ is the maximum value. The scaled distance is defined by,
\begin{equation}
    \mathcal{D}(\Phi_{p}^{i}) = \frac{d_{max} - d(\Phi_{p}^{i})}{d_{max} - d_{min}}
    \label{eq:21}
\end{equation}
where $d(\Phi_{p}^{i})$ is the distance from the sample point $i$ to the evaluated points $k$. The variables $d_{min}$ and $d_{max}$ are the minimum and the maximum distance, respectively. The best point is chosen as a candidate, measured by $\mathcal{F}(\Phi_{p}^{i})$ in eq.~\eqref{eq:17}. The objective function is tabulated for the best point \cite{wang2014}. The surrogate is updated using this value and searched again. The algorithm steps are presented in \textbf{Algorithm 1}.
\begin{center}
\begin{tabular}[c]{p{0.25cm} p{7.70cm}}
\hline
\multicolumn{2}{p{7.75cm}}{\textbf{Algorithm 1:} Surrogate optimization} \\
\hline
  & \textbf{Start} \\
1. & m independent points $\Phi_{p}^{1}$, $\Phi_{p}^{1}$, $\hdots$, $\Phi_{p}^{m}$ are created with the bounds provided by the user.\\
2. & Maximum number of iterations $\leftarrow$ $K$ \\
3. & \textbf{For} $1$ $<$ $k$ $<$ $K$ \textbf{do}\\
4. & ~~~Run the simulation model and compare it with the measured data to obtain the function values $\mathcal{G}^{1}, \mathcal{G}^{2}, \hdots, \mathcal{G}^{n}$ \\
5. & ~~~Tabulate the RBF interpolator $s$ defined in eq.~\eqref{eq:17} for the $n$ points $(\Phi_{p}^{1}, \mathcal{G}^{1})$, $(\Phi_{p}^{2}, \mathcal{G}^{2})$, $\hdots$, $(\Phi_{p}^{n}, \mathcal{G}^{n})$ \\
6. & ~~~Obtain the best point tabulated by eq.~\eqref{eq:19}, \eqref{eq:20}, and \eqref{eq:21} \\
7. & ~~~Evaluate the objective function defined in eq.~\eqref{eq:14} for the best point obtained in \textbf{Step 6} \\
8. & ~~~Update the surrogate with the value tabulated in \textbf{Step 7} \\
9. & \textbf{End} \\
10. & Return the best value for parameter set $\Phi_{p}$. \\
\hline
\end{tabular}
\end{center}

\section{Experimental Set-up and Data Acquisition}\label{section4}
This section provides a brief description about the Alaska Center for Energy and Power Energy Technology Facility (ETF) setup, components used for the experimental validation of this work, and data acquisition.

\subsection{Diesel Electric Generator Experimental Setup}
\label{subsec:expsetup}
The Alaska Center for Energy and Power ETF is a 480 VAC single bus microgrid that integrates a diesel electric generator, PV emulator, wind turbine emulator, battery energy storage, and two load banks. The components used in this work are the diesel electric generator and two load banks. The diesel electric generator is a CAT C-15 with a nameplate capacity of 400 kVA. The diesel electric generator is controlled using a Woodward EasyGen 3200 controller. The load banks are 313 kVA resistive/inductive load banks that can be controlled in 5 kW/3.75 kVAR load steps. 

In this work, load step changes were introduced in the system using the load banks and the diesel electric generator response was recorded. Initially, the load banks were set to 80 kW and 0 kVAR. The load was then increased to 240 kW and 160 kVAR. The diesel electric generator model developed in this work was simulated in the MATLAB\textsuperscript{\textregistered}/Simulink\textsuperscript{\textregistered} environment, with the same load step change of 80 kW and 0 kVAR to 240 kW and 160 kVAR. The simulation results were then compared to the experimental test data. The CAT C-15 test data acquisition is explained in the following subsection.

\subsection{CAT C15 Diesel Electric Generator Data Acquisition}
The CAT C15 diesel electric generator data was collected at the ETF using a Yokogawa DL850 oscilloscope with a 50~kHz sampling rate. All three phases of instantaneous voltage ($V_{an}$, $V_{bn}$, $V_{cn}$) and line currents ($I_a$, $I_b$, $I_c$) were collected. The root mean square (RMS) voltage ($V$), real power ($P$), and reactive power ($Q$) were calculated using the measured instantaneous voltage and current data. The $P$ and $Q$ values were calculated using eq.~\eqref{eq:12} and \eqref{eq:13}, where $V_1$ and $I_1$ are the positive-sequence component of the measured instantaneous voltage and current, respectively. The frequency ($f$) was calculated from the instantaneous voltage using a 3-phase phase-locked loop (PLL).
\begin{equation}
    P = 3\times \frac{|V_1|}{\sqrt{2}}\times \frac{|I_1|}{\sqrt{2}} \times \cos{(\phi)}
    \label{eq:12}
\end{equation}
\begin{equation}
    Q = 3\times \frac{|V_1|}{\sqrt{2}}\times \frac{|I_1|}{\sqrt{2}} \times \sin{(\phi)}
    \label{eq:13}
\end{equation}

\section{Simulation Results and Discussion}\label{section5}
\begin{figure}[!htp]
    \centering
    \includegraphics[width=0.65\columnwidth]{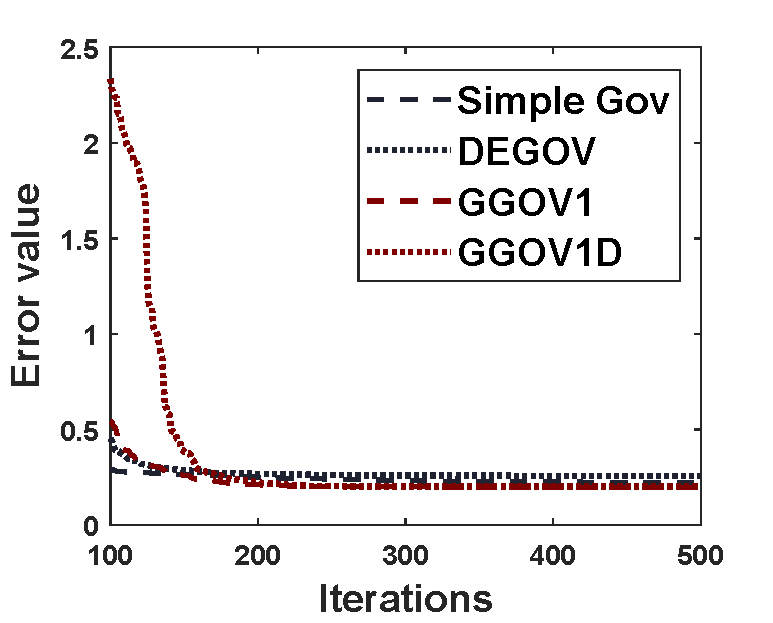}
    \caption{Error values for diesel electric generator model during load change.}
    \label{nrmseconvergence}
\end{figure}

\begin{figure}[!htp]
    \centering
    \includegraphics[width=0.95\columnwidth]{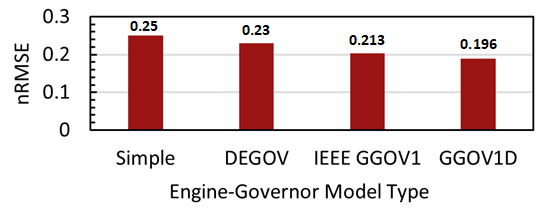}
    \caption{Comparison of diesel electric generator nRMSE values for different engine-governor models.}
    \label{nrmsecomp}
\end{figure}

The parameters for the diesel electric generator model developed in this work were estimated using the surrogate optimization. The surrogate optimization was simulated using MATLAB\textsuperscript{\textregistered} Global Optimization Toolbox \cite{matlabsurrogate} for 500 iterations with all the error weights $w_p$, $w_q$, $w_v$, and $w_f$ equal to 1. The optimization algorithm was also simulated with other error weights, but all weights equal to 1 resulted in the best results for all the responses, and due to brevity, only the results for all error weights equal to 1 are presented in this work. The objective of surrogate optimization was to minimize the cumulative nRMSE of the diesel electric generator's $P$, $Q$, $V$, and $f$ response during the load change, defined in eq.~\eqref{eq:13}.  The cumulative nRMSE values of the $P$, $Q$, $V$, and $f$ response of the diesel electric generator model with the four different engine-governor models is shown in Figs.~\ref{nrmseconvergence} and \ref{nrmsecomp}. Due to the initial nRMSE errors being very large, the nRMSE values in Fig.~\ref{nrmseconvergence} are shown from iteration 100 to 500 as opposed to 0 to 500. The nRMSE results show that the proposed GGOV1D engine-governor model outperforms other models. 

The diesel electric generator model parameters estimated using the surrogate optimization are shown in TABLE~\ref{ModelTable}. Initially, the parameters were estimated for the diesel electric generator with a simple engine-governor model. The machine and DC4B excitation system parameters were not optimized while estimating the parameters for other engine-governor models. The parameter ranges used to optimize the machine and DC4B excitation systems were obtained from the literature \cite{DC4B,Idlbi,Janssen}. The parameter range for the simple engine-governor model was determined based on the parameter values presented in \cite{stavrakakis1995, kariniotakispart2}. The parameter range for the DEGOV engine-governor model is selected from \cite{neplangov} and for IEEE GGOV1 from \cite{nercturbinegov}. The parameter range for GGOV1D model is based on the DEGOV and IEEE GGOV1 engine-governor models.
\begin{table}[!htp]
\renewcommand{\arraystretch}{1.3}
\caption{diesel electric generator parameter values}
\label{ModelTable}
\centering
\begin{tabular}{p{1.5cm} |p{1.25cm} p{1.35cm} p{1cm} p{1cm}}
\hline
\bfseries Parameters & \multicolumn{2}{c}{\bfseries Tabulated} & \bfseries Min & \bfseries Max\\
\hline
\multicolumn{5}{c}{\bfseries Machine} \\
\hline
H & \multicolumn{2}{c}{0.7359} & 0.3 & 0.8\\
\hline
\multicolumn{5}{c}{\bfseries Simple engine-governor} \\
\hline
$\textnormal{K}_{\textnormal{p}}$ & \multicolumn{2}{c}{13.8} & 1 & 40\\
$\textnormal{K}_{\textnormal{i}}$ & \multicolumn{2}{c}{30.9} & 4 & 40\\
$\textnormal{T}_{\textnormal{sm}}$ & \multicolumn{2}{c}{0.059} & 0.01 & 0.1\\
C & \multicolumn{2}{c}{0.97} & 0.5 & 2\\
$\textnormal{C}_{\textnormal{2}}$ & \multicolumn{2}{c}{1.04} & 0.5 & 2\\
$\textnormal{C}_{\textnormal{3}}$ & \multicolumn{2}{c}{1.79} & 0 & 2\\
\hline
\multicolumn{5}{c}{\bfseries DEGOV} \\
\hline
$\textnormal{T}_{\textnormal{1}}$ & \multicolumn{2}{c}{0.058} & 0.01 & 0.09\\
$\textnormal{T}_{\textnormal{2}}$ & \multicolumn{2}{c}{0.021} & 0.02 & 0.09\\
$\textnormal{T}_{\textnormal{3}}$ & \multicolumn{2}{c}{0.49} & 0.05 & 0.65\\
$\textnormal{T}_{\textnormal{4}}$ & \multicolumn{2}{c}{0.056} & 0.002 & 0.075\\
$\textnormal{T}_{\textnormal{5}}$ & \multicolumn{2}{c}{0.0058} & 0.005 & 0.05\\
$\textnormal{T}_{\textnormal{6}}$ & \multicolumn{2}{c}{0.017} & 0.009 & 0.07\\
K & \multicolumn{2}{c}{27.2} & 6 & 50\\
\hline
\multicolumn{5}{c}{\bfseries GGOV1 and GGOV1D} \\
\hline
 & GGOV1 & GGOV1D & & \\
\hline
maxerr & 0.4 & 0.069 & 0.01 & 0.5\\
minerr & -0.48 & -0.09 & -0.5 & -0.01\\
$\textnormal{K}_{\textnormal{p}}$ & 107.75 & - & 0 & 800\\
$\textnormal{K}_{\textnormal{i}}$ & 154.05 & - & 0 & 300\\
$\textnormal{K}_{\textnormal{d}}$ & 120.92 & - & 0 & 200\\
$\textnormal{N}_{\textnormal{d}}$ & 187.9 & - & 0 & 200\\
$\textnormal{T}_{\textnormal{act}}$ & 0.79 & - & 0.01 & 0.9\\
$\textnormal{T}_{\textnormal{1}}$ & - & 0.028 & 0.01 & 0.09\\
$\textnormal{T}_{\textnormal{2}}$ & - & 0.055 & 0.02 & 0.07\\
$\textnormal{T}_{\textnormal{3}}$ & - & 0.54 & 0.1 & 0.75\\
$\textnormal{T}_{\textnormal{4}}$ & - & 0.052 & 0.01 & 0.09\\
$\textnormal{T}_{\textnormal{5}}$ & - & 0.01 & 0.01 & 0.05\\
$\textnormal{T}_{\textnormal{6}}$ & - & 0.042 & 0.01 & 0.09\\
K & - & 90.42 & 30 & 150\\
$\textnormal{valve}_{\textnormal{open}}$ & 68.39 & 92.86 & 10 & 125\\
$\textnormal{valve}_{\textnormal{close}}$ & -13.02 & -105.75 & -125 & -10\\
$\textnormal{K}_{\textnormal{turb}}$ & 0.35 & 0.357 & 0.35 & 0.4\\
$\textnormal{T}_{\textnormal{b}}$ & 0.78 & 0.86 & 0.1 & 0.9\\
$\textnormal{T}_{\textnormal{c}}$ & 0.15 & 0.69 & 0.1 & 0.9\\
$\textnormal{$\omega$}_{\textnormal{fnl}}$ & 0.11 & 0.11 & 0.1 & 0.14\\
\hline
\multicolumn{5}{c}{\bfseries DC4B Exciter} \\
\hline
$\textnormal{T}_{\textnormal{r}}$ & \multicolumn{2}{c}{0.062} & 0.06 & 0.08\\
$\textnormal{K}_{\textnormal{a}}$ & \multicolumn{2}{c}{335.85} & 300 & 350\\
$\textnormal{T}_{\textnormal{a}}$ & \multicolumn{2}{c}{0.0175} & 0.01 & 0.02\\
$\textnormal{Vr}_{\textnormal{min}}$ & \multicolumn{2}{c}{-10.45} & -20 & 0\\
$\textnormal{Vr}_{\textnormal{max}}$ & \multicolumn{2}{c}{14.34} & 0 & 20\\
$\textnormal{K}_{\textnormal{f}}$ & \multicolumn{2}{c}{0.014} & 0.01 & 0.02\\
$\textnormal{T}_{\textnormal{f}}$ & \multicolumn{2}{c}{1.56} & 1.25 & 1.75\\
$\textnormal{K}_{\textnormal{e}}$ & \multicolumn{2}{c}{0.61} & 0.5 & 1\\
$\textnormal{T}_{\textnormal{e}}$ & \multicolumn{2}{c}{0.042} & 0.01 & 0.05\\
$\textnormal{K}_{\textnormal{p}}$ & \multicolumn{2}{c}{434.48} & 400 & 600\\
$\textnormal{K}_{\textnormal{i}}$ & \multicolumn{2}{c}{441.2} & 350 & 450\\
$\textnormal{K}_{\textnormal{d}}$ & \multicolumn{2}{c}{221.19} & 150 & 250\\
$\textnormal{N}_{\textnormal{d}}$ & \multicolumn{2}{c}{36.42} & 20 & 40\\
$\textnormal{K}_{\textnormal{g}}$ & \multicolumn{2}{c}{0.97} & 0 & 1\\
\hline
\end{tabular}
\end{table}

\begin{figure}[!htp]
    \centering
    \includegraphics[width=\columnwidth]{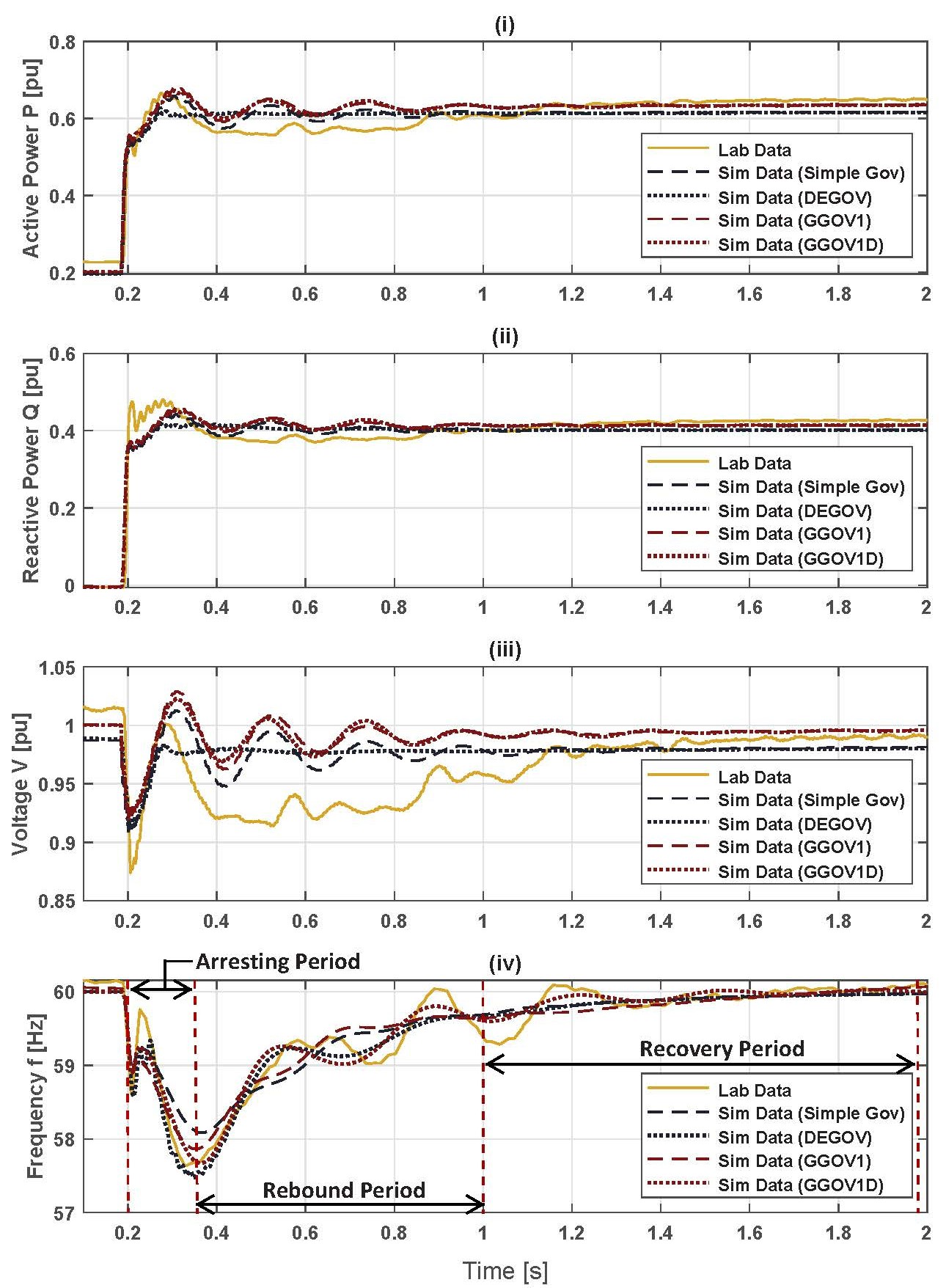}
    \caption{(i) Active power response of diesel electric generator during load change, (ii) Reactive power response of diesel electric generator during load change, (iii) Voltage response of diesel electric generator during load change, and (iv) Frequency response of diesel electric generator during load change.}
    \label{simresults}
\end{figure}

The active and reactive power, voltage, and frequency responses of the diesel electric generator model simulated using the estimated parameters shown in TABLE~\ref{ModelTable} as compared to the experimental results from the CAT C15 diesel electric generator are shown in Fig.~\ref{simresults}. From the results in Figs.~\ref{simresults}(i) and \ref{simresults}(ii), the $P$ and $Q$ responses of the diesel electric generator with GGOV1 and GGOV1D engine-governor model are a better fit to the laboratory results as compared to the diesel electric generator model with DEGOV and simple engine-governor models. The initial overshoot in the $P$ and $Q$ response during the load change using the GGOV1 and GGOV1D engine-governor model closely emulates the lab data as compared to the response using the DEGOV and simple engine-governor model. The initial voltage drop during the load change is approximately similar to the laboratory results for all the models and can be observed in Fig.~\ref{simresults}(iii). The voltage is a RMS voltage measured using a variable frequency RMS block developed in Simulink\textsuperscript{\textregistered}. The results presented in Fig.~\ref{simresults}(iv) depict that the frequency nadir of the diesel electric generator with the GGOV1D engine-governor model more closely matches the laboratory results as compared to the response using other engine-governor models. Also, the simple and IEEE GGOV1 engine-governor models have a first order actuator and hence do not capture frequency transients in the rebound period. It can also be observed from Fig.~\ref{simresults}(iv) that GGOV1D outperforms other engine-governor models in capturing frequency transients after the load change in the arresting and rebound periods. The frequency measurements were obtained using the 3-phase PLL block in Simulink\textsuperscript{\textregistered}.

\begin{figure}[!htp]
    \centering
    \includegraphics[width=0.95\columnwidth]{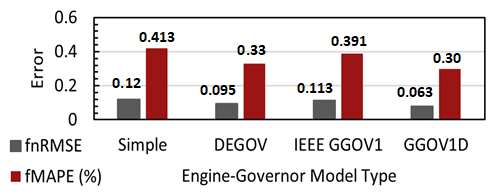}
    \caption{Comparison of diesel electric generator frequency response nRMSE and MAPE during the arresting and rebound period for different engine-governor models.}
    \label{ferrorcomp}
\end{figure}

The results in Fig.~\ref{nrmsecomp} show that the overall reduction in the nRMSE of diesel electric generator response during the load change with the GGOV1D engine-governor is approximately 27.5$\%$, 17.35$\%$, and 8.67$\%$ as compared to simple, DEGOV, and IEEE GGOV1 engine-governor models, respectively. This reduction in the nRMSE of the response is calculated across the entire 4 second response from the load change to the system reaching a steady state. The nRMSE and mean absolute percentage error (MAPE) of the frequency response during the arresting and rebound periods is shown in Fig.~\ref{ferrorcomp}. The results in Fig.~\ref{ferrorcomp} show that the diesel electric generator model with a MAPE of the GGOV1D frequency response during the arresting and rebound periods is decreased by 37.67$\%$, 10$\%$, and 30.33$\%$ as compared to simple, DEGOV, and IEEE GGOV1 engine-governor models, respectively. This improvement in the frequency response during the arresting and rebound periods using GGOV1D is important while designing the control system that can coordinate with the FFR of an inverter-based resource controller and provide frequency support to restore system frequency.

\section{Conclusion}\label{section6}
A high-fidelity diesel electric generator model was developed and tested in this work with four different engine-governor models to emulate an actual diesel electric generator response to load change. The exciter, V/Hz limiter, machine, and engine-governor parameters were tuned based on the technical specifications of the CAT C-15 diesel electric generator. The results are presented for the load step change from 80 kW and 0 kVAR to 240 kW and 160 kVAR. The results showed that the active power, reactive power, and voltage response error of the diesel electric generator with GGOV1D engine-governor model is reduced compared to the diesel electric generator with other engine-governor models. Hence it is concluded that the proposed GGOV1D engine-governor model outperforms other models. It is also concluded from the results that there is a significant improvement in the frequency response of the diesel electric generator after the load change during the arresting and rebound periods using the proposed GGOV1D model.

In the future, this high fidelity model for diesel electric generators will be used to develop control systems to provide frequency support in hybrid microgrids employing diesel electric generators and IBR  since coordination of PFR and FFR is vital as penetration of IBR increases in diesel electric generator based microgrids.


\section*{Acknowledgment}
The contributions to this research work were achieved in part through the U.S. Department of Energy Office of Science EPSCoR Program. The views expressed in the article do not necessarily represent the views of the DOE or the U.S. Government. The U.S. Government retains and the publisher, by accepting the article for publication, acknowledges that the U.S. Government retains a nonexclusive, paid-up, irrevocable, worldwide license to publish or reproduce the published form of this work or allow others to do so, for the U.S. Government purposes.

The authors would like to thank David Light at Alaska Center for Energy and Power (ACEP) for his valuable comments.

\bibliographystyle{IEEEtran}
\bibliography{main}

\vspace{-1em}
\begin{IEEEbiography}[{\includegraphics[width=1in,height=1.25in,clip,keepaspectratio]{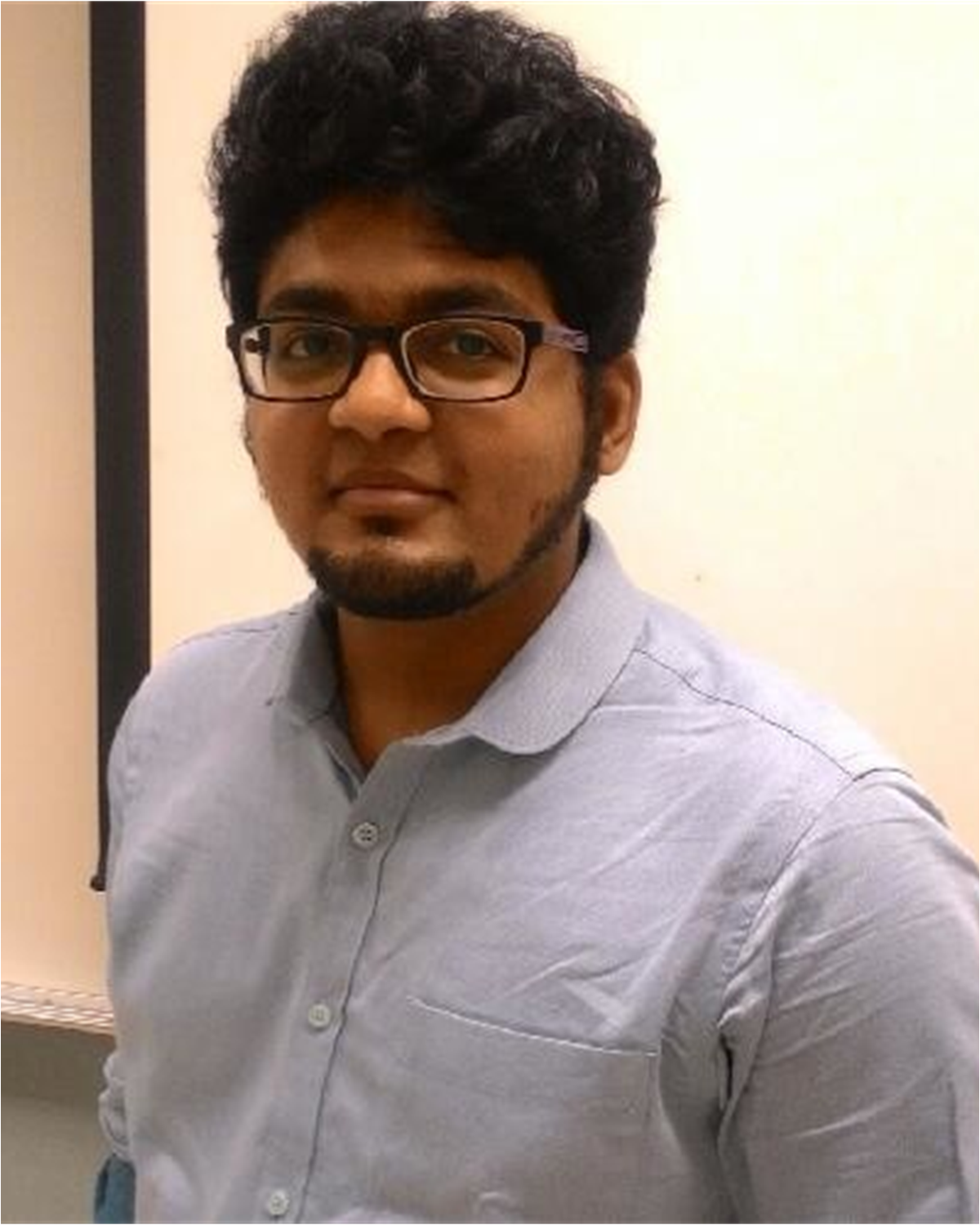}}]{Chinmay Shah} (S'15) received his B.Tech in instrumentation and control engineering from Nirma University, India in 2012 and M.S. in electrical engineering from University of Houston, TX, USA in 2017. Currently, he is pursuing a Ph.D. in electrical engineering at the University of Alaska Fairbanks.

He worked as an instrumentation and control engineer at Dodsal Engineering and Construction, Dubai, UAE, from 2012 to 2014. He also worked as a research intern at National Renewable Energy Laboratory in Summer 2019. Currently, he works as a research assistant at the Alaska Center for Energy and Power (ACEP). His research interest includes modeling and control of power electronic converter-based DERs, distributed optimization, distributed controls for the power grid, power system resiliency and reliability.
\end{IEEEbiography}

\vspace{-2em}
\begin{IEEEbiography}[{\includegraphics[width=1in,height=1.25in,clip,keepaspectratio]{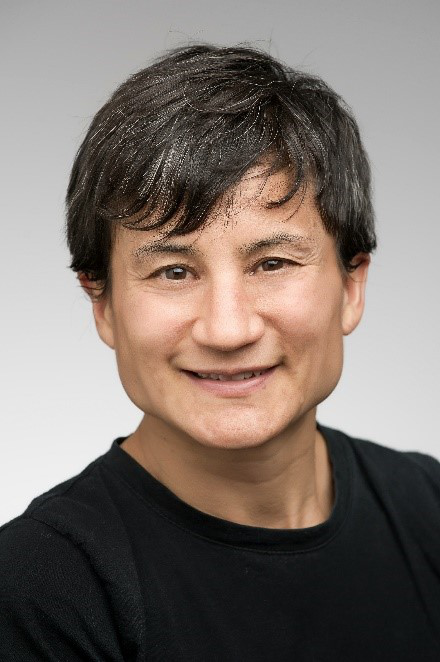}}]{Mariko Shirazi} received the B.S. degree in Mechanical Engineering from the University of Alaska, Fairbanks (UAF), in 1996 and the M.S. and Ph.D. degrees in Electrical Engineering from the University of Colorado, Boulder, in 2007 and 2009 respectively.  

Mariko was an Engineer at the National Renewable Energy Laboratory for 15 years, working on early efforts to integrate wind into village power systems, and later on power electronics design for microgrid applications.  
Mariko currently serves as the President’s Professor of Energy for the University of Alaska, where she is interested in bridging power electronics and power systems research to understand the performance of converter-dominated microgrids.
\end{IEEEbiography}

\vspace{-2em}
\begin{IEEEbiography}[{\includegraphics[width=1in,height=1.25in,clip,keepaspectratio]{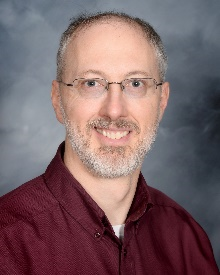}}]{Richard W. Wies} (S’92–M’99–SM’17) received his B.S., M.S., and Ph.D. degrees in electrical engineering from the University of Wyoming, Laramie, WY, USA, in 1992, 1995, and 1999, respectively.

Since 1999, he has been at the University of Alaska, Fairbanks, AK, USA, where he is currently a professor in the Electrical and Computer Engineering Department, with a concentration in electric power systems. He leads research focused on the engineering challenges of renewable energy integration in remote islanded microgrids in collaboration with the Alaska Center for Energy and Power.

He is a Licensed Professional Engineer in the State of Alaska.
\end{IEEEbiography}

\vspace{-1em}
\begin{IEEEbiography}[{\includegraphics[width=1in,height=1.25in,clip,keepaspectratio]{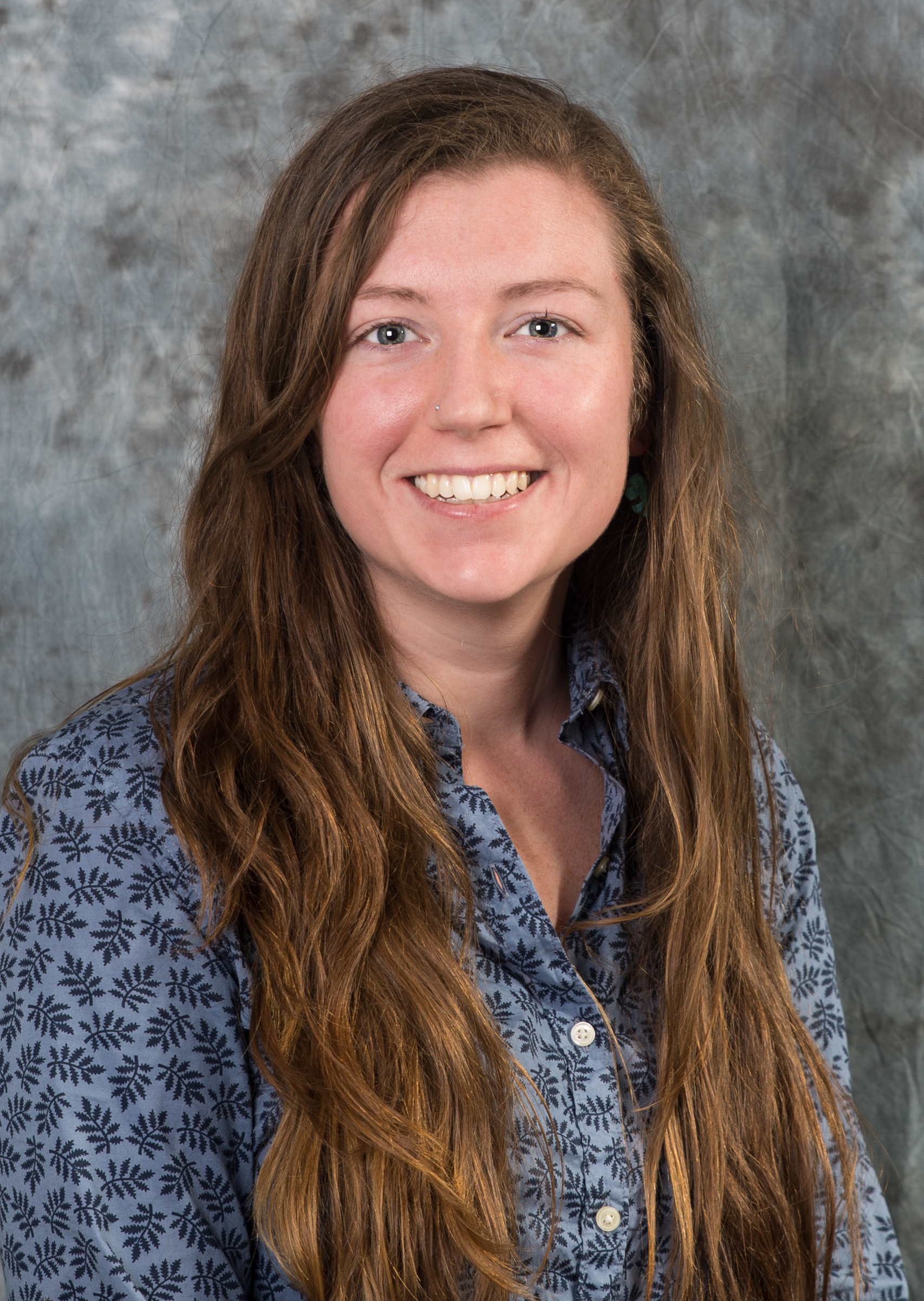}}]{Phylicia Cicilio}
(S'15-M'20) is a Research Assistant Professor at the Alaska Center for Energy and Power at the University of Alaska, Fairbanks. She received the B.S.~degree in chemical engineering in 2013 from the University of New Hampshire, Durham, NH, USA. She received the M.S.~and Ph.D.~degrees in electrical and computer engineering in 2017 and 2020 from Oregon State University, Corvallis, OR, USA. 

Her research interests include power system reliability and dynamic power system modeling particularly of loads, inverter-based resources, and distributed energy resources.
\end{IEEEbiography}

\begin{IEEEbiography}[{\includegraphics[width=1in,height=1.25in,clip,keepaspectratio]{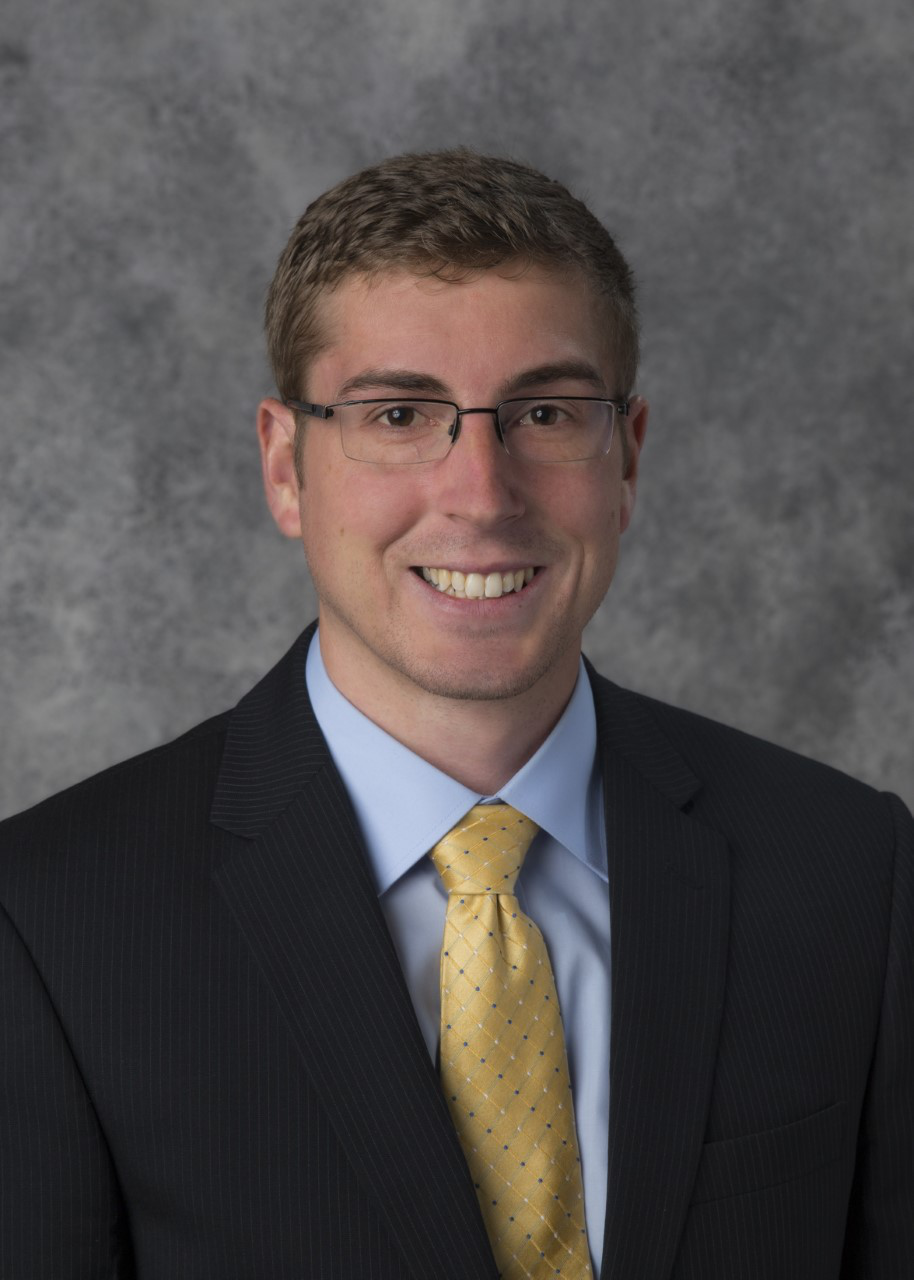}}]{Timothy M. Hansen} (S'07--SM'11--M'15--SM'20) received the B.S. in computer engineering degree from the Milwaukee School of Engineering, Milwaukee, WI, USA, in 2011, and the Ph.D. in electrical engineering degree from Colorado State University, Fort Collins, CO, USA, in 2015. 

He is currently an Assistant Professor with the Electrical Engineering and Computer Science Department, South Dakota State University, Brookings, SD, USA. His research interests are in the areas of optimization, high-performance computing, and electricity market applications to sustainable power and energy systems, low-inertia power systems, smart cities, and cyber-physical-social systems. 

Dr. Hansen is also an active member in ACM SIGHPC. He was the recipient of the 2019 IEEE-HKN C. Holmes MacDonald Outstanding Teaching Award, and was the inaugural recipient of the Milwaukee School of Engineering Graduate of the Last Decade award in 2020. Within IEEE he is the IEEE Siouxland Section Chair (since 2019) and is active within the IEEE PES Power Engineering Education Committee, currently serving as the Awards Subcommittee Chair.
\end{IEEEbiography}

\begin{IEEEbiography}[{\includegraphics[width=1in,height=1.25in,clip,keepaspectratio]{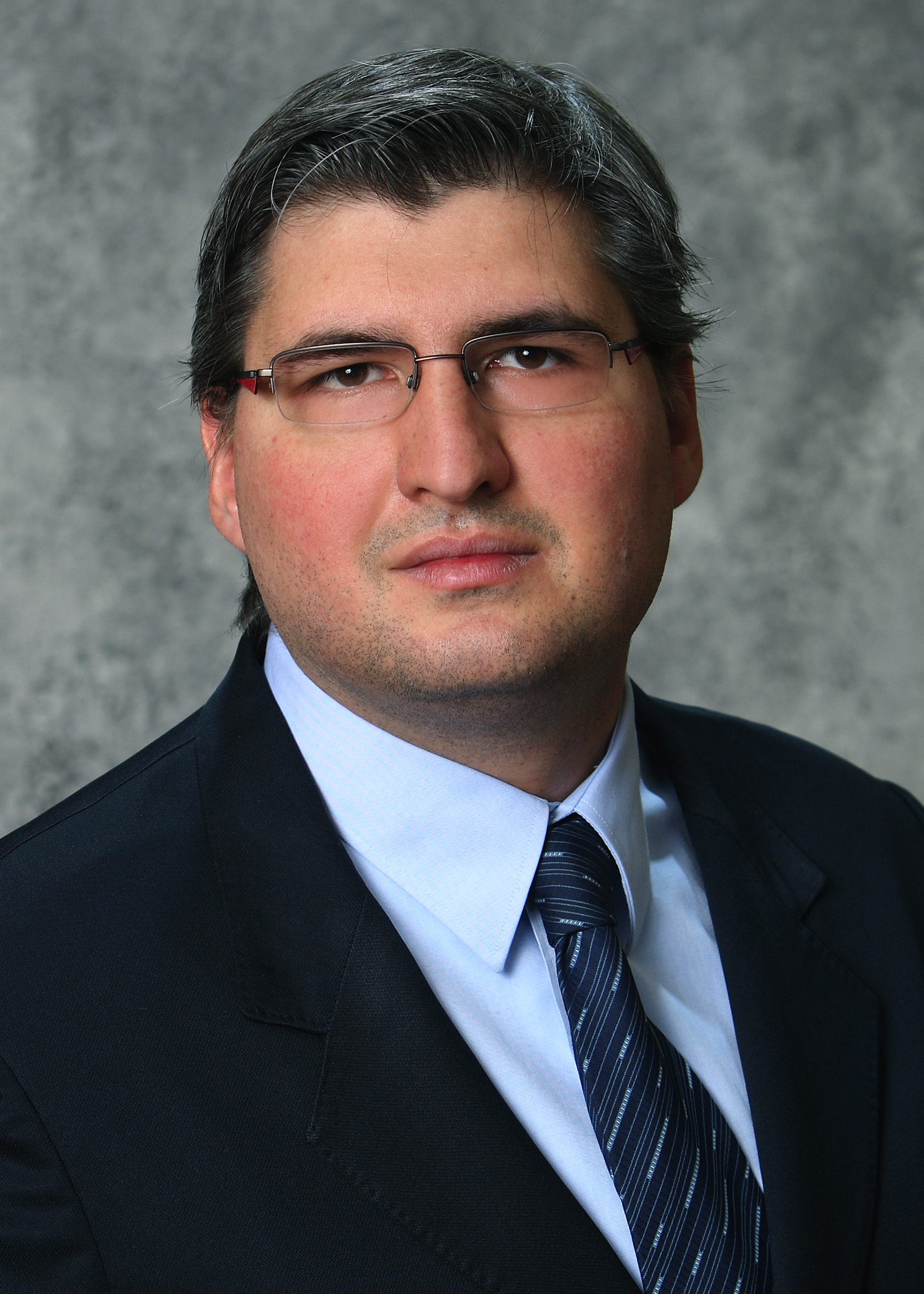}}]{Reinaldo Tonkoski}(S'04--M'11--SM'18) is the Harold C. Hohbach Endowed Professor in the Electrical Engineering and Computer Science Department at South Dakota State University, USA and a Visiting Professor at Sandia National Laboratories. He received his B.A.Sc. degree in Control and Automation Engineering, in 2004 and his M.Sc. in Electrical Engineering in 2006 from PUC-RS (Pontifícia Universidade Católica do RS), Brazil, and, his Ph.D. in 2011 from Concordia University, Canada. 

Dr. Tonkoski has authored over one hundred technical publications in peer reviewed journals and conferences and is currently an Editor of IEEE Transactions on Sustainable Energy, IEEE Access and IEEE Systems Journal. 
His research interests include grid integration of sustainable energy technologies, energy management, power electronics and control systems.
\end{IEEEbiography}

\end{document}